\documentstyle[epsfig]{aipproc}
\newcommand{\beq}{\begin{equation}}
\newcommand{\eeq}{\end{equation}}
\newcommand{\bea}{\begin{eqnarray}}
\newcommand{\eea}{\end{eqnarray}}
\newcommand{\apj}{{\it Astrophys. J.} }
\newcommand{\prl}{{\it Phys. Rev. Lett.} }
\newcommand{\rem}[1]{ }
\begin{document}
\title{Generation of Magnetic Fields and Jitter Radiation in GRBs. 
I. Kinetic Theory}

\author{Mikhail V. Medvedev }
\address{Canadian Institute for Theoretical Astrophysics,
University of Toronto,\\ Toronto, Ontario, M5S 3H8, Canada\\
Harvard-Smithsonian Center for Astrophysics, 60 Garden Street,
Cambridge, MA 02138}

%\lefthead{LEFT head}
%\righthead{RIGHT head}
\maketitle

\begin{abstract}
We present a theory of generation of strong (sub-equipartition)
magnetic fields in relativistic collisionless GRB shocks. These fields 
produced by the kinetic two-stream instability are tangled on very 
small spatial scales. This has a clear signature in the otherwise 
synchrotron(-self-Compton) $\gamma$-ray spectrum. Second, we present an 
analytical theory of jitter radiation, which is emitted when the 
correlation length of the magnetic field is smaller then the gyration 
(Larmor) radius of the accelerated electrons. We demonstrate that the 
spectral power $P(\nu)$ for pure jitter radiation is well-described 
by a sharply broken power-law: $P(\nu)\propto\nu^1$ for $\nu<\nu_j$
and $P(\nu)\propto\nu^{-(p-1)/2}$ for $\nu>\nu_j$, where $p$ is the
electron power-law index and $\nu_j$ is the jitter break, which is 
independent of the magnetic field strength and depends on the shock
energetics and kinematics. Here we mostly focus on the first problem.
The radiation theory and comparison with observations will be discussed
in the forthcoming publications.
\rem{ Finally, we present a composite 
jitter+synchrotron model of GRB emission which is capable of resolving 
several puzzles of GRB spectra, e.g., (i) the violation of the 
``line of death'' for the synchrotron shock model, (ii) sharp spectral
breaks, inconsistent with synchrotron spectrum, and (iii) multiple
spectral components seen in some bursts (good examples are
GRB910503, GRB910402, etc.). We stress that simultaneous detection of 
both spectral components opens a way to a precise diagnostics of the 
conditions in GRB shocks.}
\end{abstract}

\section*{Introduction}

There is currently no satisfactory explanation for the origin of
strong magnetic fields required in GRB shocks.  Compression of the ISM
magnetic field in external shocks yields a field amplitude $B\sim\gamma
B_{\rm ISM}\sim 10^{-4} (\gamma/10^2)~{\rm gauss}$, which is too weak
and can account only for $\epsilon_B=(B/B_{eq})^2\le 10^{-11}$
(here $\gamma$ is the Lorentz factor of the outflow).
Neither a turbulent magnetic dynamo, nor the magnetic shearing 
(Balbus-Hawley) instability, nor any other MHD process can be 
so efficient to produce the required strong fields. In principle, some
magnetic flux might originate at the GRB progenitor and be carried by the
outflowing fireball plasma. Because of flux freezing, the field amplitude 
would decrease as the fireball expands. In this case, only a progenitor with 
a rather strong magnetic field $\sim10^{16}~{\rm gauss}$ might produce 
sufficiently strong fields during the GRB emission. However, since the 
field amplitude scales as $B\propto R^{-4/3}$, even a highly 
magnetized plasma at $R\sim 10^7~{\rm cm}$ would possess only a negligible 
field amplitude of $\sim10^{-2}~{\rm gauss}$, or $\epsilon_{B}\le 10^{-7}$, at 
a radius of $R\ge 10^{16}~{\rm cm}$, where the afterglow radiation is emitted.
Here we discuss how strong magnetic fields are generated by the kinetic
relativistic two-stream instability \cite{ML99} and consider their properties. 
We postpone the major discussion on the jitter radiation theory \cite{M00}
to a forthcoming publication.

\section*{The Two-Stream Instability}

The non-relativistic instability was first discovered by Weibel \cite{W59}. 
It has been used by Moiseev and Sagdeev \cite{MS63} to develop a theory of 
collisionless non-relativistic shocks in the interplanetary space.

\begin{figure}[b!] % fig 1
\centerline{\epsfig{file=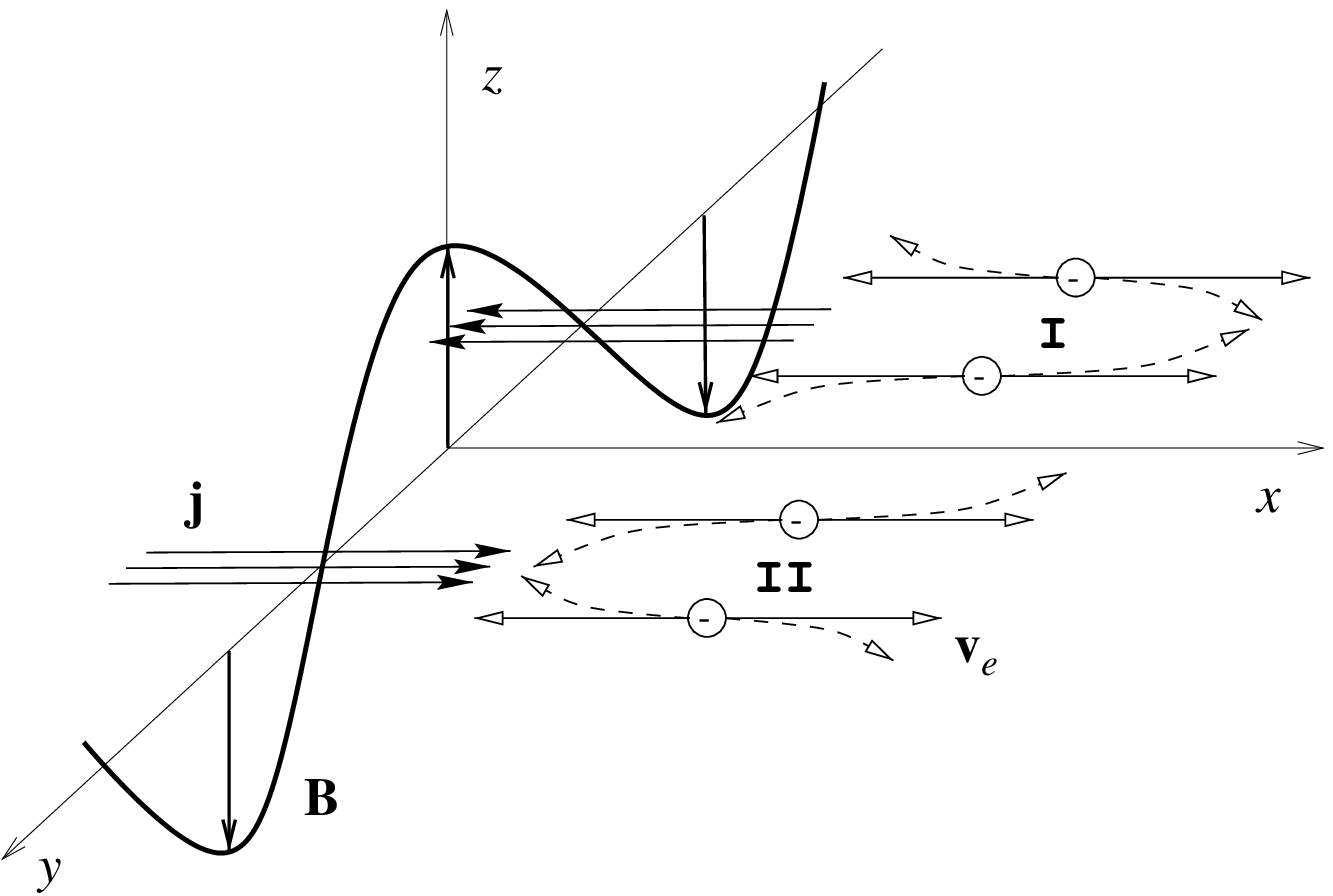,height=1.9in}~~~~
\epsfig{file=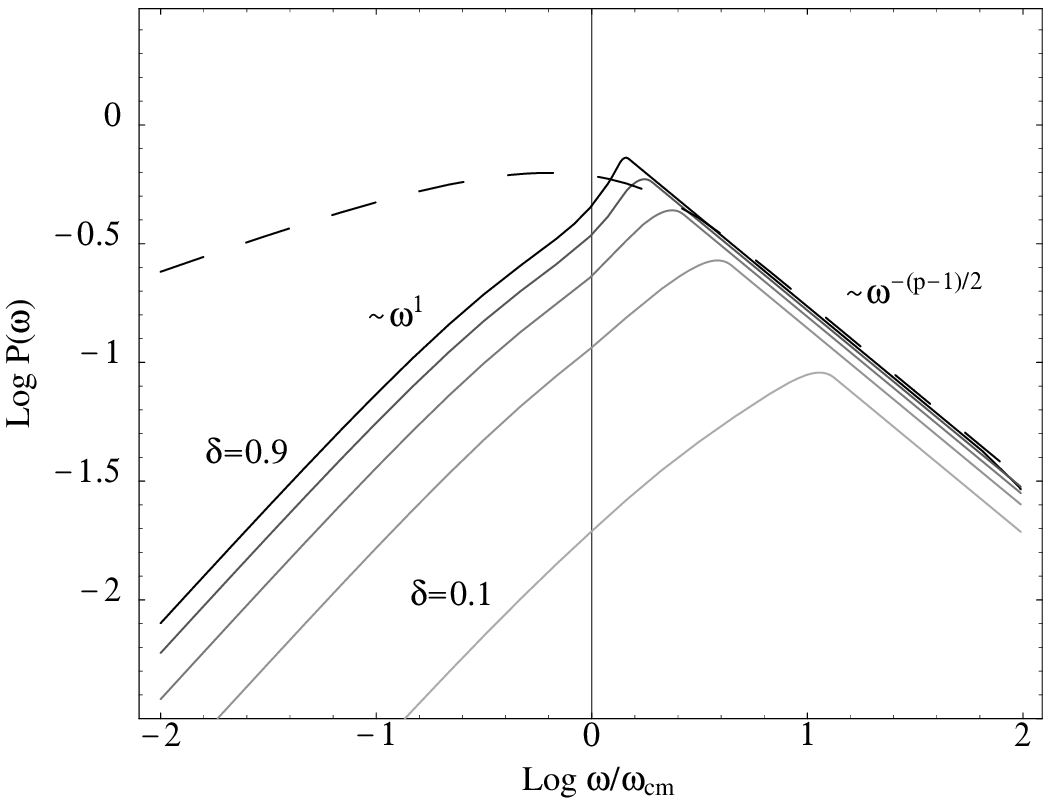,height=1.9in}}%,width=3.5in}}
\vspace{10pt}
\caption{({\it a}) --- Illustration of the instability. A magnetic field
perturbation deflects electron motion along the $x$-axis, and results in
current sheets ($j$) of opposite signs in regions I and II, which in turn
amplify the perturbation. The amplified field lies in the plane
perpendicular to the original electron motion.
({\it b}) --- Typical jitter spectra from small-scale magnetic fields
({\em dashed} curve is synchrotron, for comparison).}
\label{fig1}
\end{figure}

Let us consider, for simplicity, the dynamics of the electrons only, and
assume that the protons are at rest and provide global charge neutrality.
The electrons are assumed to move along the $x$-axis (as illustrated in
Fig.\ \ref{fig1}a) with a velocity ${\bf v}=\pm {\bf \hat x}v_x$ and equal
particle fluxes in opposite directions along the $x$-axis (so that the net
current is zero). Next, we add an infinitesimal magnetic field fluctuation,
${\bf B}= {\bf\hat z}B_z\cos(ky)$. The Lorentz force, $-e{{{\bf v}\over
c}\times{\bf B}}$, deflects the electron trajectories as shown by the
dashed lines in Fig.\ \ref{fig1}a. As a result, the electrons moving to the
right will concentrate in layer I, and those moving to the left -- in
layer II. Thus, current sheaths form which appear to {\em increase} the
initial magnetic field fluctuation. The growth rate is $\Gamma=\omega_{\rm
p}v_y/c$, where $\omega_{\rm p}^2=(4\pi e^2n/m)$ is the non-relativistic
plasma frequency \cite{Fried59}. Similar considerations imply that
perpendicular electron motions along $y$-axis, result in oppositely
directed currents which suppress the instability. The particle motions
along ${\bf\hat z}$ are insignificant as they are unaffected by the
magnetic field. Thus, the instability is driven by the anisotropy of a
particle distribution function and should quench for the isotropic case.

The Lorentz force deflection of particle orbits increases as the magnetic
field perturbation grows in amplitude. The amplified magnetic field is
{\em random} in the plane perpendicular to the particle motion, since it is
generated from a random seed field. Thus, the Lorentz deflections result in
a pitch angle scattering which makes the distribution function isotropic. 
The thermal energy associated with their random motions will be equal to 
their initial directed kinetic energy. This final state will bring the 
instability to saturation.

Here are the main properties of the instability and the produced 
magnetic fields:
\begin{itemize}
\item 
This instability is driven by the {\em anisotropy} of the particle
distribution function and, hence, can operate in both internal and
external shocks.
\item
The characteristic $e$-folding time in the shock frame for the instability is
$\tau\simeq{\gamma_{\rm sh}^{1/2}}/{\omega_{\rm p}}$ (where 
$\gamma_{\rm sh}$ is the shock Lorentz factor) which is $\sim10^{-7}~{\rm s}$ 
for internal shocks and $10^{-4}~{\rm s}$ for external shocks.
This time is much shorter than the dynamical time of GRB fireballs.
\item
The characteristic coherence scale of the generated magnetic field is of
the order of the relativistic skin depth
$\lambda\simeq2^{1/4}{c\bar\gamma^{1/2}}/{\omega_{\rm p}}$
(where $\bar\gamma$ is the mean thermal Lorentz factor of particles), i.e. 
$\sim10^3~{\rm cm}$ for internal shocks and $\sim10^5~{\rm cm}$ for external 
shocks. This scale is much smaller than the spatial scale of the source.
\item
The generated magnetic field is randomly oriented in space, but always lies
in the plane of the shock front.
\item
The instability is powerful. It saturates only by nonlinear effects when
the magnetic field amplitude approaches equipartition with the electrons
(and possibly with the ions). Therefore 
$[{B^2/8\pi}]/[{mc^2n(\bar\gamma-1)}]=\eta\sim0.01 - 0.1\,$.
This result is in excellent agreement with direct particle simulations.
\item
The instability isotropizes and heats the electrons and protons.
\item
Random fields scatter particles over pitch-angle and, thus, provide effective 
collisions. Therefore MHD approximation works well for the shocks.
The magnetic fields communicate the momentum and pressure of the outflowing 
fireball plasma to the ambient medium and define the shock boundary.
\item 
The generated small-scale fields affect the radiation processes \cite{M00}
and produce non-synchrotron spectra of radiation, 
as shown in Fig.\ \ref{fig1}b.
\end{itemize}

\end{document}